\documentclass[manuscript]{aastex}

\def\beq{\begin{equation}}
\def\eeq{\end{equation}}
\def\Msolar{\ifmmode{\rm M}_{\mathord\odot}\else${\rm M}_{\mathord\odot}$\fi}
\providecommand{\sorthelp}[1]{}

\newcommand{\Euclid}{{\em Euclid}}

\newcommand{\PSZtwo}{{\tt PSZ2}}

\newcommand{\Ho}{H_0}
\newcommand{\OmM}{\Omega_{\rm m}}

\newcommand{\sigmaf}{\sigma_{\rm f}}
\newcommand{\Yfive}{Y_{500}}
\newcommand{\meanYfive}{\bar{Y}_{500}}

\newcommand{\Mfive}{M_{500}}
\newcommand{\thetafive}{\theta_{500}}
\newcommand{\meanthetafive}{\bar{\theta}_{500}}

\newcommand{\qcut}{q_{\rm cut}}

\newcommand{\meanqm}{\bar{q}_{\rm m}}

\slugcomment{}

\shorttitle{The Sunyaev--Zel'dovich effect}
\shortauthors{Battistelli et al.}
\usepackage{amsmath}
\usepackage{gensymb}
\usepackage{graphicx}

\begin{document}

\def\dustem{\textsc{DustEM}}
\def\Planck{\textit{Planck}}
\def\Herschel{\textit{Herschel}}
\def\Spitzer{\textit{Spitzer}}
\def\IRAS{\textit{IRAS}}

\title{Galaxy clusters as probes for cosmology and dark matter\footnote{Based  on  presentations  at  the  Fourteenth  Marcel  Grossmann  Meeting  on  General  Relativity,
Rome, July 2015.}}
\author{Elia S. Battistelli$^{1}$, Carlo Burigana$^{2,3,4}$, Paolo de Bernardis$^{1}$, Alexander A. Kirillov$^{5}$,  Gastao B. Lima Neto$^{6}$, Silvia Masi$^{1}$, Hans U. Norgaard-Nielsen$^{7}$,  Peter Ostermann$^{8}$, Matthieu Roman$^{9}$, Piero Rosati$^{3}$, Mariachiara Rossetti$^{10,11}$}

\altaffiltext{1}{Dipartimento di Fisica, Sapienza Universit\`a di Roma, P.le Aldo Moro 5, I-00185, Rome, Italy -- elia.battistelli@roma1.infn.it }
\altaffiltext{2}{INAF-IASF Bologna, Via Piero Gobetti 101, I-40129 Bologna, Italy}
\altaffiltext{3}{Dipartimento di Fisica e Scienze della Terra, Universit\`a degli Studi di Ferrara,Via Giuseppe Saragat 1, I-44122 Ferrara, Italy}
\altaffiltext{4}{INFN, Sezione di Bologna, Via Irnerio 46, I-40126, Bologna, Italy}
\altaffiltext{5}{Dubna International University of Nature, Society and Man, Dubna, 141980, Russia}
\altaffiltext{6}{Departamento de Astronomia, IAG/USP, S\~ao Paulo/SP, 05508-090, Brazil}
\altaffiltext{7}{DTU Space, Elektrovej, DK - 2800 Kgs. Lyngby, Denmark}
\altaffiltext{8}{independent-research.org, Munich, Germany}
\altaffiltext{9}{Laboratoire de Physique Nucl\'eaire et des Hautes \'Energies (LPNHE),Universit\'e Pierre et Marie Curie, Paris, France}
\altaffiltext{10}{Dipartimento di Fisica, Universit\`a degli Studi di Milano, Via Celoria 16, I-20133, Milan, Italy}
\altaffiltext{11}{INAF - IASF Milano, via Bassini 15, I-20133 Milano}

\begin{abstract}

In recent years, significant progress has been made in building
new galaxy clusters samples, at low and high redshifts, from
wide-area surveys, particularly exploiting the Sunyaev--Zel'dovich
(SZ) effect. A large effort is underway to identify and
characterize these new systems with optical/NIR and X-ray
facilities, thus opening new avenues to constraint cosmological
models using structure growth and geometrical tests. A census of
galaxy clusters sets constraints on reionization mechanisms and
epochs, which need to be reconciled with recent limits on the
reionization optical depth from cosmic microwave background (CMB)
experiments. Future advances in SZ effect measurements will
include the possibility to (unambiguously) measure directly the
kinematic SZ effect, to build an even larger catalogue of galaxy
clusters able to study the high redshift universe, and to make
(spatially-)resolved galaxy cluster maps with even spectral
capability to (spectrally-)resolve the relativistic corrections of
the SZ effect.

\end{abstract}


\section{Introduction}

Galaxy clusters are among the most studied and interesting objects
of our universe. Recent technological improvements have made
possible a massive increase of the size of galaxy cluster
catalogues and a key role has been played by the full exploitment
of the Sunyaev--Zel'dovich (SZ) effect as galaxy clusters
searcher. The SZ effect is a powerful tool to study our universe
at low and high redshift, to investigate the adiabaticity of the
universe expansion, the equation of state of dark energy, the dark
matter distribution in the universe, and the astrophysics
governing the biggest objects of our Universe. The SZ effect has
been historically described by three main contributions: the
thermal SZ effect arising from the thermal motion of ionised
medium in the intra-cluster medium (ICM), the kinematic SZ effect
arising from the peculiar motion of the galaxy clusters, and the
relativistic corrections to the SZ effect which originate when one
accounts for the relativistic temperatures characterising the ICM.

\subsection{Thermal SZ effect}

The thermal SZ effect \citep{sz70,sz72} is a spectral distortion of the cosmic
microwave background (CMB) caused by inverse Compton scattering between the CMB
photons and a hot electron gas present in the ICM in clusters of galaxies.

A complete derivation of the effect can be found in several
reviews \citep{ZS,reph95,birk99,carl02}. In the nonrelativistic limit, the spectral
behavior of the distorted spectrum can be obtained by solving the Kompaneets
equation \citep{komp1957}, which yields the occupation number $n(\nu)$ of the
radiation energy levels:
\begin{eqnarray}
\frac{\partial n}{\partial
t}=\frac{kT_{e}}{m_ec}\frac{\sigma_Tn_e}{x_{e}^2}\frac{\partial}{\partial
x_{e}}\left[x_{e}^4\left(\frac{\partial n}{\partial
x_{e}}+n+{n}^2\right)\right]\!,
\end{eqnarray}
where $n_e$ is the electron number density, $m_e$ and $T_e$ are the electron mass
and gas temperature respectively, $\sigma_T$ is the Thomson cross-section and
$x_{e}=h\nu/k_{B}T_{e}$ (not to be confused with the dimensionless frequency
$x=h\nu/k_{B}T_{\rm CMB}$).

Under the assumption that $T_{e}\gg T_{\rm CMB}$, the Kompaneets
equation admits a simple analytical solution. We define the
parameter $y$ as:
\begin{eqnarray}
y=n_{e}\sigma_{T}\frac{k_{B}T_{e}}{m_{e}c^{2}}ct,
\end{eqnarray}
which represents a dimensionless measurement of the time spent in the electron
distribution and can be written in the form:
\begin{eqnarray}
y=\int\! n_{e}\sigma_{T}\frac{k_{B}T_{e}}{m_{e}c^{2}}
dl=\tau\frac{k_{B}T_{e}}{m_{e}c^{2}},
\end{eqnarray}
where the integration is calculated along the line of sight $dl$, $\tau$ is the
optical depth and electron temperature $T_{e}$ has been assumed constant; $y$ is
known as the Comptonization paramenter. At low optical depth and low electron
temperature, for a Planckian incident radiation spectrum, we get a spectral
variation of the occupation number of the form:
\begin{eqnarray}
\Delta n(x)=xy\frac{e^{x}}{(e^{x}-1)^{2}}\left(x\cdot
\textnormal{coth}\left(\frac{x}{2}\right)-4\right)
\end{eqnarray}
and thus the specific intensity change due to thermal SZ effect is:
\begin{eqnarray}
\frac{\Delta I_{\rm TSZ}(x)}{I_{0}}=x^{4}y\frac{e^{x}}{(e^{x}-1)^{2}}\left(x\cdot
\textnormal{coth}\left(\frac{x}{2}\right)-4\right)=y g(x),\label{eq:isz}
\end{eqnarray}
where $I_{0}=\frac{2h}{c^{2}}(\frac{k_{B}T_{\rm CMB}}{h})^{3}$.

The spectral behavior of the thermal SZ effect is shown in
Fig.~1(a). From the above equations, we can extract the main
features of the thermal SZ effect in the nonrelativistic
approximation:
\begin{itemize}
\item the SZ effect amplitude depends only on $y$: it linearly depends on the cluster
electron temperature and on the optical depth $\tau$ (i.e. on the cluster pressure
integrated along the line of sight);
\item the spectral behavior is described by relatively simple analytical functions (i.e. $g(x)$);
\item we have a zero point of the effect for $x=3.83$ (i.e. $\nu=217$\,GHz)
with $\Delta I_{\rm TSZ}(x)\,{<}\,0$ for $x<3.83$ and $\Delta I_{\rm TSZ}(x)>0$ for
$x>3.83$;
\item the specific intensity is characterized by a minimum value at $x=2.26$ and a maximum at $x=6.51$;
\item since the SZ effect arises solely from interaction of CMB photons with hot cluster gas and the CMB
is a background that exists everywhere, the SZ effect does not suffer spherical
dilution or cosmological dimming as an isotropic radiator does; this results in
its redshift independence. This is valid also for its spectral behavior under the
assumption that the CMB temperature scales with the redshift as
\begin{eqnarray}
T_{\rm CMB}(z)=T_{\rm CMB}(0)(1+z).
\end{eqnarray}
\end{itemize}

\subsection{Kinematic SZ effect}\label{par:ksz}

While the thermal SZ effect is due to random thermal motion of the ICM electrons
with isotropic distribution, if the cluster has a finite peculiar velocity, one
expects an additional kinematic effect due to the electrons motion with respect to
the CMB, which causes a Doppler shift of the scattered photons. If one assumes
that thermal and the kinematic SZ effect are separable, it is forward to extract
the spectral dependence of the kinematic SZ:
\begin{eqnarray}
\frac{\Delta I_{\rm
KSZ}(x)}{I_{0}}=x^{4}\frac{e^{x}}{(e^{x}-1)^{2}}\frac{v_{p}}{c}\tau=
h(x)\frac{v_{p}}{c}\tau,
\end{eqnarray}
where $v_{p}$ is the peculiar cluster velocity along the line of sight.

The net effect observed on the CMB spectrum is thus a change of
the temperature of the Planckian spectrum, found to be higher for
negative velocities (i.e. moving toward the observer) and vice
versa (see Fig.~1). The kinematic SZ effect is a powerful method
to determine peculiar cluster velocities along the line of sight.
The spectral coincidence between the maximum intensity of the
kinematic effect and the zero point of the thermal effect allows,
in principle, to distinguish between them. The spectral behavior
of the kinematic SZ effect, is anyway identical to that of the CMB
anisotropies: this results in an effective difficulty to
disentangle these two effects from their spectral behavior (see
Fig.~1(b)).

\begin{figure}
\begin{tabular}{ccc}
\includegraphics[width=5.3cm]{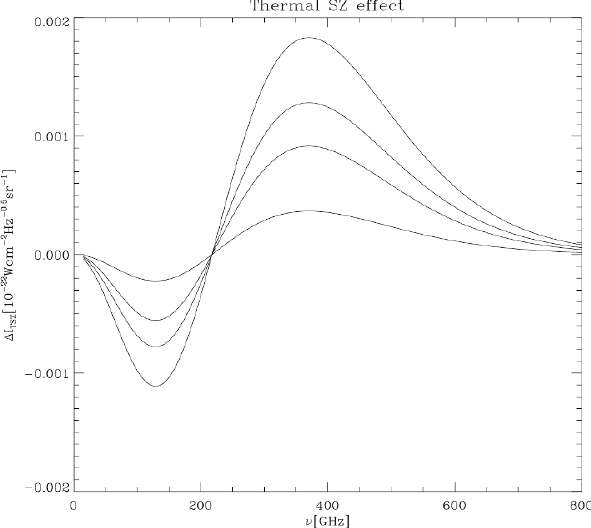}
\includegraphics[width=5.3cm]{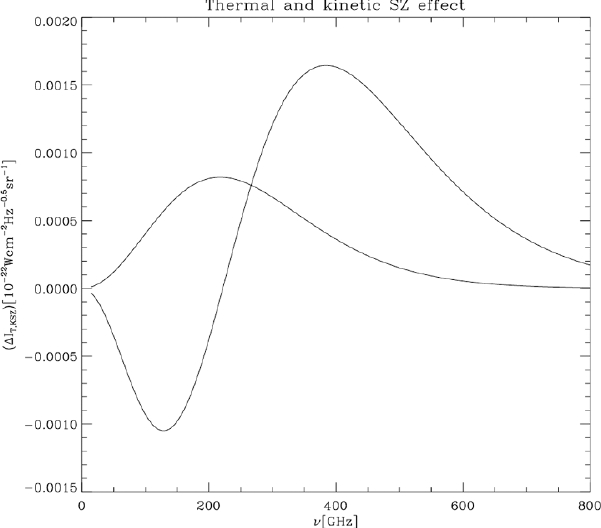}
\includegraphics[width=5.3cm]{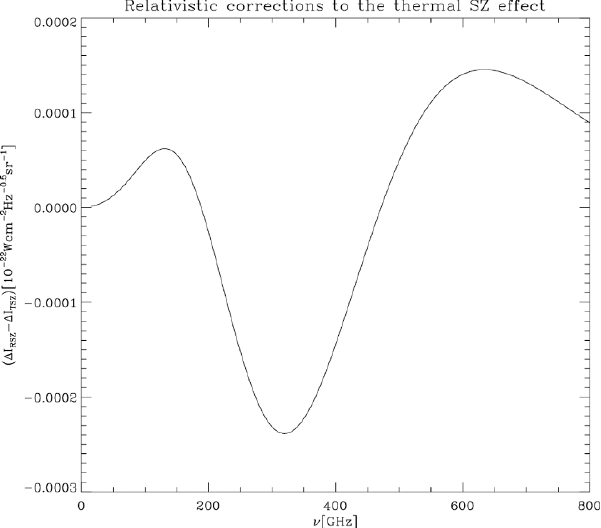}
\end{tabular}
\caption{Left: thermal SZ effect for $y=10^{-4}$, $7 \times 10^{-5}$, $5 \times 10^{-5}$, $2 \times 10^{-5}$. Centre: thermal and kinematic SZ effect for  $y=10^{-4}$ , $v_{p}/c=10^{-3}$ and $T_{e}=8.2$keV. The kinematic curve has been multiplied by a factor -10. Right: relativistic corrections to the SZ effect up to the $5^{th}$ order in $\Theta_{e}$ for $y=10^{-4}$, and $T_{e}=8.2$keV.}
\label{fig:sz}
\end{figure}

\subsection{Relativistic corrections}\label{par:relSZ}

ICM electrons with temperatures from $\simeq\!3$\,keV are characterized by near
relativistic velocities. In order to perform a $more$ $exact$ calculation of the
SZ effect, one needs to take into account relativistic corrections to it. Taking
into account these effects drives to nonnegligible corrections of the order of a
few percent of the thermal SZ effect itself.

Different methods have been proposed in order to deal with this problem: a
numerical approach consists in performing Monte Carlo simulations of interactions
between electrons and photons in a fully relativistic regime. The analytical
approach allows to explicit the relativistic correction in terms of powers of the
electron cluster temperature.

The inadequacy of the Kompaneets equation had already been
considered early in 1979 \citep{wright} and in 1981 \citep{Fabbri81} when it was extended the formalism to the case
of little number of scatterings. An explicit treatment for typical
cluster temperatures has been presented by \cite{reph95b}, who has stressed, in agreement with \cite{itoh98} the shift of the zero point frequency $x_{0}$
of the thermal SZ effect: $x_{0}$ is pushed to higher
values with increasing electron temperature
$\Theta_{e}=\frac{k_{B}T_{e}}{m_{e}c^{2}}$. \cite{chal98} focused of the corrections in the
Rayleigh--Jeans part of the spectrum while a second-order
approximation which takes into account additional corrections
related with the peculiar cluster velocity has been found by \cite{Saz99}. An analytical fitting formula for
the relativistic corrected thermal SZ effect, $\frac{\Delta
I_{{\rm T}_{R}{\rm SZ}}(x)}{I_{0}}$ up to the fifth-order in
$\Theta_{e}$ has been reported by \cite{itoh98}{:}

\begin{equation}
\frac{\Delta I_{T_{R}SZ}(x)}{I_{0}}=yx^{4}\frac{e^{x}}{(e^{x}-1)^{2}}(Y_{0}+\Theta_{e}Y_{1}+\Theta_{e}^{2}Y_{2}+\Theta_{e}^{3}Y_{3}+ \Theta_{e}^{4}Y_{4})=yg_{rel}(x,T_{e}) 
\end{equation} 
where $Y_{n}$ are a set of functions depending on the dimensionless frequency
only. The spectral behavior of the relativistic corrections to the thermal SZ
effect is shown in Fig.~1(c), once subtracted by the uncorrected thermal SZ
spectrum.

\cite{noza98} have added to the above expression the terms
originated by the effect of the peculiar cluster velocity, while \cite{itoh01} have extended the relativistic correction up to the
seventh-order in power of $\Theta_{e}$ and \cite{shim04} have
found an accurate expression of the cross over frequency up to the fourth-order in
$\Theta_{e}$ considering also the dependence on $\tau$.

In the relativistic corrected treatment the spectral behavior of the thermal SZ
effect is thus dependent on $T_{e}$ and the one of the kinematic SZ effect depends
on both $T_{e}$ and $v_{p}$. Multifrequency observations of the SZ effect are
necessary to disentangle the different effects. In principle, one could use the
kinematic SZ and the relativistic corrections of the SZ effect to infer the
properties of the galaxy clusters.

\clearpage

\section{The SZ Effect as Cosmological Probe}

\subsection{Absolute CMB temperature at galaxy cluster redshift$\/:$\\ The Melchiorri--Rephaeli method}

The possibility to use multifrequency observations of the SZ effect to constrain
the absolute temperature of the CMB at the redshift of the observed galaxy cluster
was already suggested long ago by \cite{Fabbri78} and by
\cite{reph80} who have proposed the method independently. Francesco
Melchiorri and Yoel Rephaeli have first applied and led these analyses since
then \citep{mel,rep}, demonstrating the method capability when the first
multifrequency millimetric SZ measurements have been available. We here propose to
name this method after them: the {\em Melchiorri--Rephaeli} method (M--R method).
The proposed analysis is based on the steep frequency dependence of the change in
the CMB spectral intensity, $\Delta I_{\rm SZ}$, due to the effect, and the weak
dependence of ratios $\Delta I_{\rm SZ}(\nu_{i})/\Delta I_{\rm SZ}(\nu_{j})$ of
intensity changes measured at two frequencies ($\nu_{i},\, \nu_{j}$) on properties
of the cluster. Because of this, and the fact that, in the Standard Model (SM) the
effect is redshift-independent, SZ measurements have the potential of yielding
much more precise values of $T_{\rm CMB}(z)$ that can be obtained from atomic and
molecular lines detection. With the improved capability of reasonably precise
spectral measurements of the SZ effect, the M--R method can now be used to measure
$T_{\rm CMB}(z)$ in a wide range of redshifts.

The M--R method has been applied for the first time to SZ measurements performed by MITO
experiment on Abell 1656 \citep{dep02} and by SuZIE experiment on Abell 2163 \citep{hol97}
clusters of galaxies \citep{Battistelli02} and several other SZ measurements have been used
since then \citep{hor05,lamagna07,luzzi09,esz, mul13, hur14,saro14, planck2, dem15, luzzi15}.
This measurement can effectively yield very precise measurements of $T_{\rm CMB}(z)$
which will tightly constrain alternative models for the functional scaling of the CMB
temperature with redshift, and thereby provide a strong test of nonstandard cosmological
models. \cite{lima00} assume an ``adiabatic photon creation'' which
takes place for instance because of a continuous decaying vacuum energy density or some
alternative mechanisms of quantum gravitational origin. A functional form which seems
also to be of some theoretical interest is described by \cite{losecco01}\/:
\begin{equation} 
T_{CMB}(z)=T_{CMB}(0)(1+z)^{1-\beta}. 
\end{equation}  
Obviously, in the SM $\beta=0$. More generally, deviations from
isotropy and homogeneity or models in which ratios of some of the
fundamental constants vary over cosmological time are also of
considerable interest. A precise determination of $T_{\rm CMB}(z)$
would allow to put constraints on the variation of the fundamental
constants over cosmological time giving thus important information
about the previously cited alternative theories or other exotic
cosmologies. So far the best limits for arising from the M--R
method are obtained by the {\em Planck}\footnote{{\it Planck} is a
project of the European Space Agency~--- ESA~--- with instruments
provided by two scientific Consortia funded by ESA member states
(in particular, the lead countries: France and Italy) with
contributions from NASA (USA), and telescope reflectors provided
in a collaboration between ESA and a scientific Consortium led and
funded by Denmark.\vadjust{\vspace*{-3pt}}} data to the percentage
level in $\beta$ \citep{luzzi15,avg15} (see Fig.~2). It is worth
noting that further limits on these alternative cosmologies will
be achieved through CMB spectral distortions
measurements \citep{chluba14} but it is also clear that a new
phenomena/physics has to be hypothesized in case $\beta\neq 0$ is
found.

\begin{figure}
\includegraphics[width=16cm]{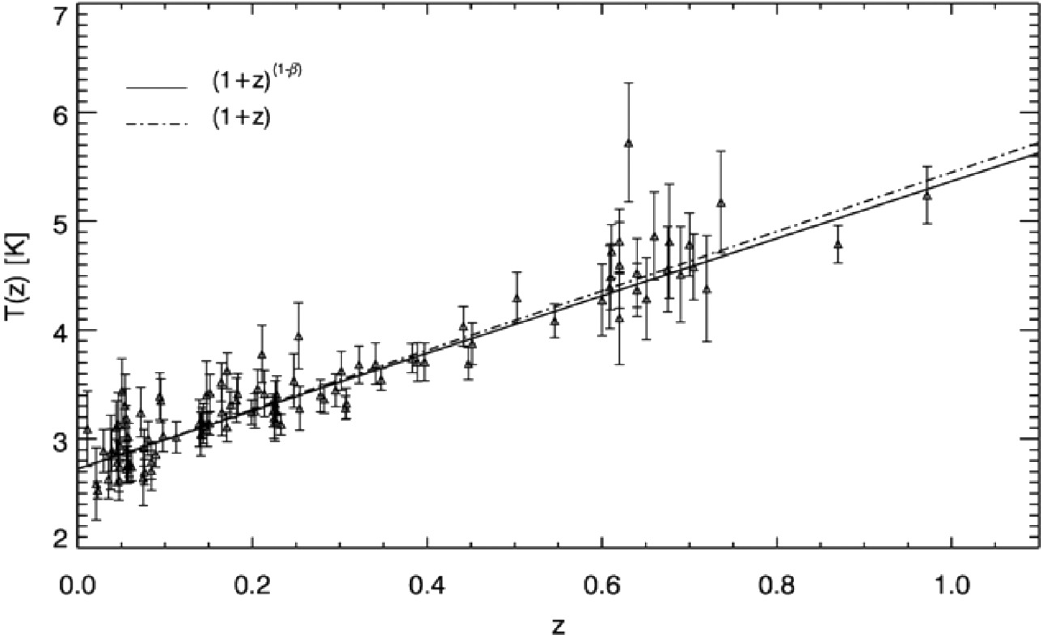}
\caption{Measurements of the CMB temperature as a function of redshift for 104 {\em Planck} selected clusters. The solid line is the best fit to to the scaling
$T_{CMB}(z)=T_{CMB}(0)(1+z)^{1-\beta}$. The dot-dashed line is the standard scaling, with $\beta =0$. Picture taken from  \cite{luzzi15}, see there for details (courtesy G. Luzzi).}
\label{fig:tvsz}
\end{figure}

\subsection{SZ as baryonic dark matter searcher}

The SZ effect also represents a unique probe to study the baryon distribution in
our universe enabling us to solve, at least in part, the apparent discrepancy
between the observed baryon distribution in the local universe (by HI absorption,
gas and stars in galaxy clusters and X-ray emission,
$\Omega_{b}\simeq\!0.021)$ \citep{fuk98} as compared to the one predicted by
nucleosynthesis, by Ly-alpha forest and by CMB power spectrum observations
($\Omega_{b}\simeq\!0.049$) \citep{planck2}. A diffuse baryonic dark matter is in
fact predicted by simulations \citep{cen99,dav01,shu12} in the form of warm-hot
intergalactic medium (WHIM). Detections have been performed in the ultraviolet and
soft X-ray bands \citep{tri08,nic05,buo09} but most of them are ambiguous or
controversial and they lack of a confirmation by later studies \citep{yao12}. The
presence of WHIM could be detectable via SZ effect, especially in super-clusters
of galaxies (SCG) where the intra-supercluster medium (ISCM) may be sufficiently
dense, hot or characterized by large line of sights, to be seen through the SZ
effect even if no clusters of galaxies are present in that region. Search for this
kind of signature have been performed with different approaches, either via direct
measurements \citep{gen05,bat06,pipVIII} or cross correlating CMB maps with matter
tracing templates \citep{sua13}. The first clear detection of filamentary structures
was performed by the {\em Planck} Collaboration in the intermediate
release \citep{pipVIII} for the merging cluster pair A399--A401, and three more in
2015 results \citep{plaXXII}. \cite{van14} reported a
positive correlation between maps reconstructed from the Canadian France Hawaii
Telescope Lensing Survey and a thermal SZ map constructed from {\em Planck} data.
They interpreted this correlation as arising from WHIM at $z \simeq\!0.4$ with
temperature in the range $10^{5}$--$10^{6}$\,K.

This research field has enormously benefited by the high sensitivity, wide
spectral coverture, and improved angular resolution obtained with the {\em Planck}
satellite. Finer angular resolution, improved sensitivity, and deeper control of
foreground removal, together with improved cross-correlation techniques, will
allow to shed a definitive light on this still open problem.

\subsubsection{The collapsing supercluster ${\rm SC}0028-0005$ at redshift $0.22$}

Besides being possible locations to search for baryonic dark matter, superclusters of
galaxies are of tremendous astrophysical interest to understand the $\Lambda$CDM standard
structures formation scenario and multi-frequencies studies have to be performed for a
deep understanding of the underlying physics. Superclusters of galaxies are in fact
collapsing objects, barely out of the linear evolution phase of the primordial density
fluctuations. The low density contrast between supercluster and the field, just above
unity, makes their detection burdensome. Moreover, given their large size their dynamical
analysis is difficult and only a few superclusters have been thoroughly studied up to now
(see e.g. Refs.~33, 6 and~79).

The supercluster SC0028-0005 \citep{omi16} at redshift 0.22 was used to investigate
the structure and dynamics of a super-cluster, which was first detected by
Basilakos \citep{Basilakos03}, who used a percolation algorithm applied to the
cluster catalogue of Goto \citep{Goto02}. Using the Sloan Digital Sky Survey Data
Release 10, all member galaxies within 1.2 degrees from the supercluster center
were selected.

The clusters and groups comprising the supercluster were identified by determining
the highest density peaks in the projected 2D density distribution, with the
constraint that these structures should be dominated by a red population of
galaxies. Six clusters (or groups) members were then used as probes of the cluster
dynamics.

A multi-band observation done with MegaCam/CFHT was also used to derive the mass
distribution inside the supercluster based on weak-lensing effect. The mean seeing
was $0.5^{\prime\prime}$ and the data processing was done by the Terapix team.

The 3D spatial distribution was determined by measuring the
fundamental plane of elliptical galaxies. Once the distance was
known, the peculiar line-of-sight velocity could be retrieved. The
supercluster mass distribution was estimated by applying the
spherical collapse model within the context of a $\Lambda$CDM
cosmology. The equation of motion of a spherical shell together
with the background Friedman--Lema\^{\i}tre equation were solved
simultaneously.

The dynamical analysis suggests that SC0028 is indeed a collapsing
supercluster, supporting the $\Lambda$CDM standard formation
scenario of these structures. The mass within $r=10$\,Mpc was
estimated to lie between 4 and $16 \times 10^{15} M_\odot$. The
farthest detected members of the supercluster imply that within
${\sim}60$\,Mpc the density contrast is $\delta \sim\!3$ with
respect to the critical density at the supercluster redshift, so
that the total mass is $\sim\!4.6$--$16 \times 10^{17} M_\odot$,
most of which is in the form of galaxy group sized or smaller
substructures. More details can be found in \cite{OMill15}.

\subsection{SZ versus X-ray emission}
As it is well known, thermal SZ and X-ray emissions sample the
same ICM with different properties. This is due to the fact that
the SZ effect is proportional to the pressure integrated along the
line of sight, $\Delta I_{\rm SZ} \propto \int n_{e}T_{e} dl $
while the X-ray emission depends on the cluster characteristics in
different way: $S_{x} \propto \int n_{e}^{2} \Lambda_{X} dl$,
where $\Lambda_{X}$ is the X-ray cooling function, weakly
depending on $T_{e}$. Joint analysis of the SZ and X-ray data
allows to study the pressure profiles of galaxy clusters and brake
degeneracy between the ICM gas density and the cluster angular
diameter distance.

This analysis has been historically used for the determination of
the Hubble constant $H_{0}$ \citep{carl02,bat03}. This has the great
advantage to be independent on extragalactic standard candles; on
the other hand, the method is found to be affected by calibration
uncertainties \citep{ree10} and by the accurate modeling of the
cluster density and temperature profile \citep{bon11}. Early
measurements assumed the gas to be isothermal and tended to
underestimate $H_{0}$. So does a departure from the sphericity of
the galaxy cluster if the cluster is elongated along the line of
sight (and vice versa). More recent studies account for radial
temperature profile of $T_{e}$ but direct temperature measurements
are available only up to about one third of the viral radius from
spatially resolved X-ray measurements and rely on simulations for
a careful determination of the temperature
profile \citep{kit14}. The fact that many galaxy clusters are often
aspherical, far from being relaxed and well approximated by a
standard profile increases the scatter in these measurements. In
addition, offsets between the X-ray emission and the SZ peaks in
merging clusters have cosmological implications \citep{zha14}.
Selection bias may also apply as elongated-along-the-line-of-sight
clusters are statistically more prone to be observed. Clumpiness
in the gas density tend to overestimate $H_{0}$ while
inhomogeneities of the temperature underestimates it. All this,
not to account for point source contamination in the SZ
measurements.

Similar analysis can be performed for the determination of the gas mass fraction
of galaxy clusters \citep{kit14}. Reversing the argument, one can try to infer
galaxy cluster properties and any departure from the distance duality relation
using the same SZ and X-ray observations assuming cosmological parameters
including $H_{0}$ from other measurements. 

\section{SZ as Clusters Finder}

Thanks to its redshift independence, the SZ effect
can actually be used to discover galaxy clusters in blind
large-area microwave maps enabling, through optical follow-up
observations, an effective count of galaxy clusters as a function
of the redshift. Galaxy cluster counts are a standard cosmological
tool that has found powerful applications. The spectral distortion
imprinted on CMB by the SZ effect can be easily recognized with a
multi-frequency high-sensitivity instrument like {\it Planck} or
with improved angular resolution surveys as in the case of the
Atacama Cosmology Telescope (ACT) or the South Pole Telescope
(SPT). Recent studies have demonstrated the SZ effect to be a very
efficient method to detect clusters \citep{act,spt}, and to provide
a complementary view to X-rays to study ICM properties, especially
in the faint outer regions \citep{pipV, pipX, eckert_sz}. The sample
of galaxy clusters observed through the SZ effect has been
enlarged by more than an order of magnitude in the last
decade \citep{act,spt,PSZ2}.

The ability to find a galaxy cluster in a microwave/millimetric
map is mainly related to the cluster flux and mass. A typical
galaxy cluster at redshift $\simeq\!1$ has a size in the sky of
the order of $2$~arcmin and an adequate angular resolution has to
be achieved in order to efficiently detect distant clusters. The
ACT \citep{hin10} and the SPT \citep{sta09} have been designed to
achieve such an angular resolution. ACT and SPT have been able to
find clusters down to $M_{500} \simeq\!1.5 \cdot 10^{14}$ solar
masses up to $z \simeq\!1.5$ (where $M_{500}$ is the cluster mass
enclosed in $R_{500}$ which corresponds to a radius within which
the average matter density is $500$ times the critical matter
density of the universe). Despite {\em Planck} experiment is
characterized by a more moderate angular resolution, it has
performed a full sky survey with much wider frequency coverage.
This makes {\em Planck} more stable for observing nearby
clusters\footnote{Part of this paper is based largely on the
products available at the ESA {\it Planck} Legacy Archive and
publicly available publications by ESA and the {\it Planck}
Collaboration, for what concerns the related aspects. Any material
presented here that is not already described in {\it Planck}
Collaboration papers represents the views of the authors and not
necessarily those of the {\it Planck} Collaboration.} with respect
to ACT and SPT. The SPT has been the first experiment to blindly
detect a galaxy cluster \citep{sta09} and, since then, 516 optically
confirmed clusters have been found by the SPT \citep{spt}, 91 by the
ACT \citep{act} and 1094 by {\em Planck} \citep{PSZ2}. See Fig.~3 for
a resume of the current SZ discovered clusters with their mass as
a function of the redshift.

The abundance of clusters and its evolution are sensitive to the
cosmic matter density, $\OmM$, and the present amplitude of
density fluctuations, characterized by $\sigma_8$, the RMS linear
overdensity in spheres of radius $8h^{-1}$\,Mpc, as well as the
nature of dark energy. CMB anisotropies reflect the density
perturbation power spectrum at the time of recombination. A
comparison of the two can be used to test the evolution of density
perturbations from recombination until today, enabling us to look
for possible extensions to the base $\Lambda$CDM model.

One of the importance of studying the SZ clusters number counts,
is that it is mainly independent from other cosmological probes
like baryonic acoustic oscillations (BAO), Super Novae Ia
measurements and even primary CMB anisotropies measurements. A
fundamental point is however related to the ability to determine
the galaxy clusters mass to which any use of the number counts as
cosmological probe is sensitive to. This is particularly critical
for high redshift clusters for which spatially resolved X-ray
imaging and gravitational lensing estimates are more and more
difficult. In this sense, we can rely on the scaling {\em Y-M}
relations between observed SZ distortion parameter and cluster
mass. Despite being still uncertain, these scaling relations are
already being used and will make the SZ even a more powerful tool
in future surveys.

\begin{figure}
\centering
\includegraphics[width=16cm]{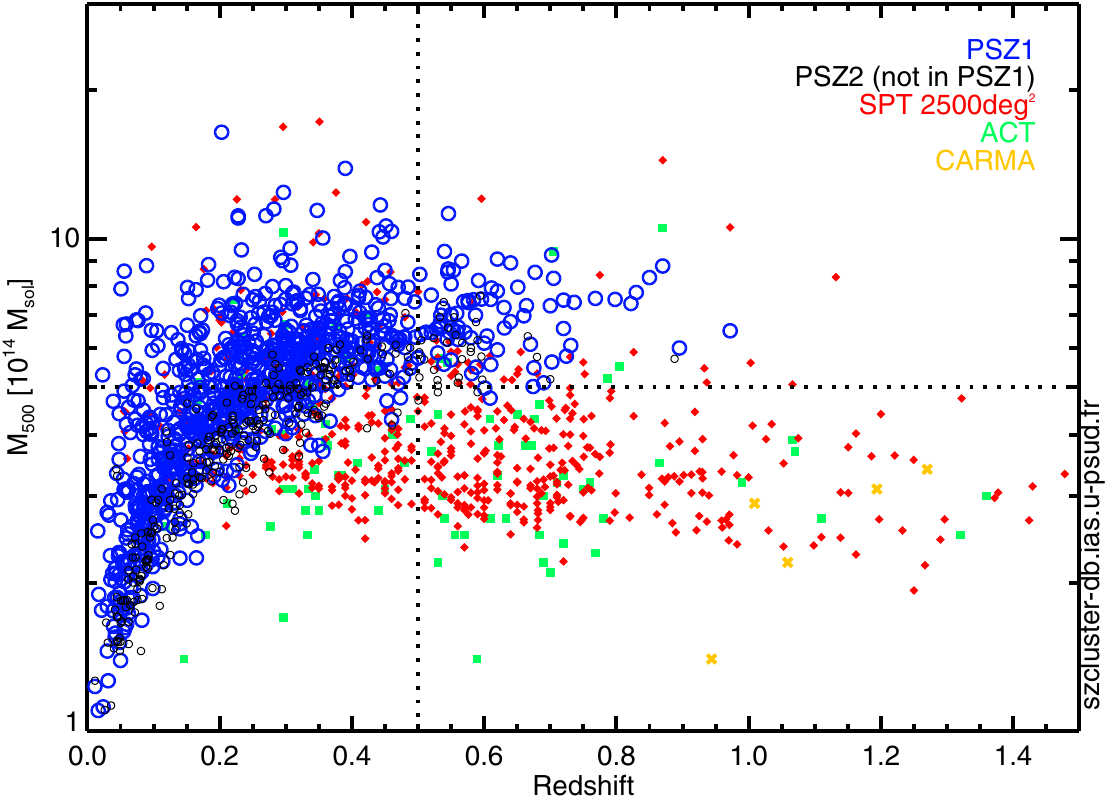}
\caption{Clusters detected by ACT, SPT and {\em Planck} (as well as Carma \citep{bon12}) as a function of redshift. This compilation was downloaded from http://szcluster-db.ias.u-psud.fr. }
\label{fig:counts}
\end{figure}

Comparison, in the 2013 {\em Planck} release, between estimates of
$\sigma_8$ and $\OmM$ from clusters number counts and from primary
CMB anisotropies found inconsistent results. Despite being
sensibly relaxed in the 2015 {\em Planck} release (see
Sec.~\ref{pla}), the origin of this tension is not completely
sorted out. Besides the possibility of an incomplete instrumental
calibration, this tension could be reconciled assuming
underestimate of the true clusters masses or missing an important
fraction of clusters, which would have driven to an
underestimation of $\sigma_8$ in the cluster counts analysis, or
assuming a variation of initial perturbation spectrum and/or
suppression of density fluctuations at small scales (like those
possibly produced by massive neutrinos), which would have driven
to an overestimation of $\sigma_8$ in the primary CMB analysis.

It is worth noting though that, as shown by Hasselfield {\it
et~al.} \citep{act}, different {\em Y-M} scaling relations have a
large impact on the $\Omega_{m}$, $\sigma_{8}$ determination.
Reversing the argument one could use the SZ counts with CMB
measurements to infer the {\em Y-M} scaling relation
normalization.

\subsection{The second {\it Planck} catalogue of SZ sources}
In this section, we present the Second {\it Planck} Catalogue of
Sunyaev-Zel'dovich sources (PSZ2 hereafter), which is part of the
2015 {\it Planck} full-mission data release \citep{PSZ2}. The PSZ2
is the third catalogue of galaxy clusters detected by {\it
Planck}, after the ESZ \citep{esz}, containing 189 detections from
the first months of data, and the PSZ1, \citep{psz1} containing
1227 sources detected by {\it Planck} in its nominal mission
($15.5$ months). The PSZ2 contains 1653 clusters candidates
detected from the full mission data (29 months with both LFI and
HFI instruments) with a signal-to-noise ratio (SNR) larger than
$4.5$ on $83.6\%$ of the sky (excluding the Galactic plane and
known point sources). The catalogue is publicly available in the
{\em Planck} Legacy Archive (PLA, http:/\!/pla.esac.esa.int/pla/),
along with the survey completeness function (we refer to the
PSZ2 \citep{PSZ2} and PSZ1 \citep{psz1} papers for details on data
analysis and for a description of the products in the archive).

The SZ observable is the integrated Comptonization parameter
$Y_{\rm SZ}$, which is measured in the {\it Planck} catalogues on
a scale of $5R_{500}$. Since the SZ parameter is strongly
degenerate with the cluster size, the {\it Planck} Collaboration
provided for each detection the two-dimensional posterior
distribution for $Y_{5R500}$ and the scale radius $\theta_s$. The
value $Y_{5R500}$ provided in the catalogues is the expected value
of the marginal distribution and is shown to be unbiased at a few
percent level. This measurement could be possibly rescaled to the
more physically interesting Comptonization parameter on the
$R_{500}$ scale by assuming a pressure profile. However, a
posterior analysis has shown that this procedure returns strongly
biased results. Therefore, to estimate $Y_{500}$ (and thus
$M_{500}$) from {\it Planck} data, it is necessary to break the
size-flux degeneracy by using prior information, relating
$Y_{500}$ and $\theta_{500}$ using the definition of $M_{500}$ and
the $Y_{500}-M_{500}$ scaling relation. \\
A necessary element to characterize a survey is the selection
function, both in terms of completeness (i.e. the probability that
an object with a given observable is detected in the survey) and
of statistical reliability, or purity (i.e. the probability that a
detection with a given SNR is a real cluster). The completeness is
a function of both the cluster size $\theta_{500}$ and the SZ flux
$Y_{500}$ and is available in the PLA for different SNR
thresholds. It was estimated for the PSZ2 with two independent
methods that provided consistent results: Monte Carlo injection in
simulated maps and a semi-analytical treatment (assuming Gaussian
noise). Monte Carlo simulations were also used to assess the
survey reliability: its lower limit is 83\% for a SNR threshold of
$4.5$, but it rapidly increases with SNR. The reliability is
higher if we apply a more selective mask excluding a larger region
around the Galactic plane and if we apply a higher SNR threshold,
as in the subsample used for measuring cosmological parameters
with galaxy clusters \citep{cosmoPSZ2}.

Searching for counterparts is an important part of the catalogue
construction, because it allows to validate the PSZ2 detections
and to associate redshifts (necessary for breaking the size-flux
degeneracy and thus measure the mass and the integrated signal) to
the confirmed detections. This procedure is based on the
well-validated PSZ1 catalogue and complemented with
multi-wavelengths ancillary data sets (microwave, IR, optical and
X-ray catalogues) and each possible counterpart association has
been confirmed by analysis of scaling relations. 1094 detections
in the PSZ2 were associated to a known counterpart with redshift:
the PSZ2 is thus the first SZ catalogue with more than one
thousand confirmed objects and the largest SZ-selected sample of
galaxy clusters.

In Fig.~3, in addition to the ACT, SPT and Carma clusters, the distribution of the
{\it Planck}-detected clusters with counterpart in the mass-redshift plane, along
with the expected completeness curves, is shown. The new PSZ2 detections are
mostly low mass objects close to the detection limit of the survey in all redshift
ranges. We should stress that the distribution in Fig.~3 is not fully
representative of the experiment selection, since only points with an associated
redshift could be plotted. For instance, at high redshifts ($z>0.6$) most clusters
are PSZ1 detections which were validated through dedicated follow-up. A similar
follow-up campaign on the almost 600 cluster candidates in PSZ2 would populate
further the $M-z$ plane and allow to characterize many new interesting objects.

\subsection{Science with the Planck SZ $2015$ cluster sample}\label{pla}

The first {\it Planck} 2013 cluster analysis found fewer clusters
than predicted by {\it Planck}'s base $\Lambda$CDM model,
expressed as tension between the cluster constraints on $(\OmM,
\sigma_8)$ and those from the primary CMB
anisotropies \citep{planck2013-p15}. This could reflect the need for
an extension to the base $\Lambda$CDM model, or indicate that
clusters are more massive than determined by the SZ signal-mass
scaling relation adopted\break in 2013.

The cluster mass scale is the largest source of uncertainty in
interpretation of the cluster counts. We based our first analysis
on X-ray mass proxies that rely on the assumption of hydrostatic
equilibrium. We quantified our ignorance of the true mass scale of
clusters with a mass bias parameter that was varied over the range
$[0\text{--}30]\%$, with a baseline value of 20\%, as suggested by
numerical simulations (see the Appendix of
\cite{planck2013-p15}).

New, more precise lensing mass measurements for {\it Planck} clusters have appeared since
then. In addition, we apply a novel method to measure cluster masses through lensing of
the CMB anisotropies. We incorporate these new results as prior constraints on the mass
bias in the present analysis: $1\,{-}\,b\,{=}\,0.688\,{\pm}\,0.072$ from Weighing the
Giants (WtG \citep{vonderlinden2014}), $1-b=0.780\pm0.092$ from the Canadian cluster
comparison project (CCCP \citep{hoekstra2015}) and $1/(1-b)=0.99\pm0.19$ from {\it Planck}
CMB lensing \citep{melin2014}. Two other improvements over 2013 are the use of a larger
cluster catalogue and the analysis of the counts in signal-to-noise as well as
redshift \citep{cosmoPSZ2}.

We define a cosmological sample from the general \PSZtwo\ catalogue \citep{PSZ2},
consisting of detections by the MMF3 matched filter. It is defined by a signal-to-noise
(denoted $q$) cut of $q>6$. We then apply a mask to remove regions of high dust emission
and point sources, leaving 65\% of the sky unmasked. The MMF3 sample contains 439
detections and spans masses in the range $[2\text{--}10] \times 10^{14}\,M_\odot$ and
redshifts from $z=0$ to 1.

The distribution of clusters in redshift and signal-to-noise can be written as
\begin{equation}
\label{eq:dndzdq}
\frac{dN}{dz dq} = \int d\Omega_{\rm mask} \int d\Mfive \, \frac{dN}{dz d\Mfive d\Omega}\, P[q | \meanqm(\Mfive,z,l,b)],
\end{equation}
with $dN/dz d\Mfive d\Omega$, the dark matter halo mass function times the volume
element. We adopt the mass function from \cite{tinker2008}. The quantity
$P[q|\meanqm(\Mfive,z,l,b)]$ is the distribution of $q$ given the mean
signal-to-noise value predicted by the model for a cluster of mass $\Mfive$ and
redshift $z$ located at Galactic coordinates $(l,b)$. This latter quantity is
defined as the ratio of the mean SZ signal expected of a cluster,
$\meanYfive(\Mfive,z)$, and the detection filter noise,
$\sigmaf[\meanthetafive(\Mfive,z),l,b]$. The redshift distribution of clusters
detected at $q > \qcut$ is the integral of Eq.~(\ref{eq:dndzdq}) over
signal-to-noise given $\chi(\Yfive,\thetafive,l,b)$, the survey selection function
at $q>\qcut$.

We divide the catalogue into bins of size $\Delta z =0.1$ (10 bins) and
$\Delta\log q =0.25$ (5 bins), each with an observed number $N(z_i,q_j)=N_{ij}$ of
clusters. Modeling the observed counts as independent Poisson random variables,
our log-likelihood is
\begin{equation}
\label{eq:like2d}
\ln L = \sum_{i,j}^{N_z N_q} \left[ N_{ij}\ln\bar{N}_{ij}  - \bar{N}_{ij} - \ln[N_{ij}!]  \right],
\end{equation}
where $N_z$ and $N_q$ are the total number of redshift and
signal-to-noise bins, respectively. The mean number of objects in
each bin is predicted by theory according to Eq.~(\ref{eq:dndzdq})
which depends on the cosmological (and cluster modeling)
parameters. In practice, we use a Monte Carlo Markov Chain (MCMC)
to map the likelihood surface around the maximum and establish
confidence limits.

We combine all SZ cluster counts results with Big Bang
Nucleosynthesis (BBN) constraints and BAO. We vary all six
parameters of the (flat) base $\Lambda$CDM model, except when
considering model extensions for which we include the relevant
parameters. Counts are consistent with those from 2013 and yield
comparable constraints.

The central value of the WtG mass prior lies at the extreme end of the range used
in 2013; with its uncertainty range extending even lower, the tension with primary
CMB is greatly reduced, as pointed out by \cite{vonderlinden2014}. With
similar uncertainty but a central value shifted to $1-b=0.78$, the CCCP mass prior
results in greater tension with the primary CMB. The lensing mass prior, finally,
implies little bias and hence much greater tension.

We obtain the posterior on $(1-b)$ from a joint analysis of the
MMF3 cluster counts and the CMB with the mass bias as a free
parameter. The best-fit value in this case is $(1-b)=0.58\pm
0.04$, more than 1$\sigma$ below the central WtG value. Perfect
agreement with the primary CMB would imply that clusters are even
more massive than the WtG calibration. This case most clearly
quantifies the tension between the {\it Planck} cluster counts and
primary CMB. Allowing $\tau_{\rm reion}$ to adjust offers only
minor improvement in the tension.

The SZ counts (2D likelihood with CCCP prior) only marginally break the dark
energy degeneracy when combined with the CMB, but when combined with BAO they do
yield interesting constraints that are consistent with the independent constraints
from the primary CMB combined with supernovae.

Increasing neutrino mass goes in the direction of reconciling tension between the
cluster and primary CMB constraints, but would necessarily increase tension with
some direct measurements of $\Ho$ and with BAO. The lensing power spectrum, in
favoring slightly lower $\sigma_8$, reinforces the cluster trend so that a peak
appears in the posterior for $\sum m_\nu$; it is not enough, however, to bring the
posterior on the mass bias in line with the prior. This indicates that the tension
between the cluster and primary CMB constraints is not fully resolved.

With a number of improvements and new information relative to our first study (more data,
2D likelihood), we confirm the 2013 results. We extract cosmological constraints using
three different cluster mass scales and find that the value needed for the mass bias to
bring cluster counts and CMB primary anisotropies in full agreement is larger than the
used priors, suggesting a tension at the astrophysical level or indicating a need for
extension of the baseline cosmological model. A precision of 1\% on the cluster mass,
targeted by future large lensing surveys such as \Euclid \ and LSST, would significantly
clarify the extent of any tension.

\subsection{Planck 2015 vs SZ cluster counts: an alternative cosmological model to explain the  mismatch}

In view of the {\em Planck} 2015 model prediction mismatch of SZ
cluster counts it seems intriguing to test also a recently found
mathematical chance of a universal micro-wave background composed
of redshifted radiation emitted from a homogeneously
distributed part of `dark' matter (hDM) (for details, see
\cite{Ostermann2012b}). Within a stationary Universe the
additional part might exist instead of the `dark energy' assumed
today. According to that solution the SZ effect would stay
present, though increasingly weakened with redshift. A
corresponding gradual shift of the SZ spectral profile to lower
frequencies seems ruled out at first sight. With respect to the
subtraction of unavoidable noise and various `foregrounds',
however, a definite clarification seems more difficult than
expected. In particular, the existence of the CIB as well as the
nature of inhomogeneities of the whole microwave background have
to be taken into a new consideration to clarify this question
seriously.

Inhomogeneities of the alternative CMB, for example, might also cause some
frequency shift, thereby compensating a small frequency shift in parts.
Measurements in the bands $\ge 218$\,GHz seem particularly problematic. The
results have been mostly relying on smaller frequencies before, which in view of
unknown individual masses seem to make no clear differences between both
alternatives of the thermal SZ in cluster count ranges $z<1$.

Actually it cannot be safely excluded, that the {\em Planck} 2015 model prediction
mismatch might partially arise from a correspondingly reduced signal-to-noise ratio.
Indeed, {\em Planck}'s major objectives encompassing tests for theories of inflation and
providing a direct probe into the Concordance Model's initial inhomogeneities are mainly
focused on the $\Lambda$CDM cosmology.

According to a new Stationary Universe Model
(SUM) \citep{Ostermann2012a,Ostermann2014} straightforwardly in
accordance with the SNe Ia data on universal scales
($z>0.1$) \citep{Ostermann2012b} the alternative CMB solution
requires an attenuation $1/(1+z)^{2}$ of intensity in the mm range
(due to $k = 2/R H$ in addition to the usual photon energy loss by
redshift). It is obvious that a gradual reduction of the SZ
intensity (up to e.g. 1/4 the value expected for clusters at $z =
1$) would lead to a steeper `$q=6$' slope down from the third
redshift bin in Figs.~4 or~6 of \cite{PSZ2}, while other
`free' parameters may be adjustable to match the absolute values
of the first and second redshift bin, too.

Though it seems sure that an explanation can be found within the highly adaptable
$\Lambda$CDM framework, there may be a scientific obligation to falsify the
tentative SUM approach in particular by evaluation of the {\em Planck} 2015 model
mismatch of predicted SZ cluster counts, in this case without any $\Lambda$CDM
priors temporarily.

\section{Intra Cluster Medium Physics}

The SZ effect is clearly a powerful tool also to study the astrophysics of galaxy
clusters themselves. This represents a complementary and alternative method with
respect to the classic use of X-ray observations, especially for distant clusters
and in the low density cases. Spatially resolved SZ measurements of the thermal
and dynamical state of the ICM in groups and clusters can provide, alone, radial
profiles of the electron pressure up to larger distances with respect to X-ray
observations. Once combined with X-ray observations, high angular resolution SZ
measurements promise to recover the $n_{e}$ and $T_{e}$ profiles thus providing a
unique tool to test models and simulations. This is of fundamental importance for
the comprehension of the structure formation and to study turbulent and
distant-from-equilibrium scenarios.

The physical state of the ICM depends on several studied and
simulated effects as warming up processes, radiative cooling, star
formation, cosmic rays, magnetic fields and thermal feedback. The
presence of cooling cores and the effect of merging processes
still need to be fully understood. Nonthermal electrons also
produces CMB photons scattering and cross correlation between
synchrotron and SZ emission maps would be fundamental to
understand the underlying energy distribution of nonthermal
electrons and to verify predictions \citep{col03,fer12}. Nonthermal
pressure effects seem to be of fundamental importance at large
distances from the cluster center thus making the SZ a possible
unique tool for the understanding the galaxy cluster physics.
Multi mergers complexes are excellent laboratories for high
angular resolution SZ observations in order to reveal temperature
and pressure gradients as well as dynamical state of the clusters
sub-structures. Merging activity has been detected toward a few
clusters including {\em el Gordo} (i.e. ACT-CL
J0102-4915) \citep{men12} deserving further studies. High angular
resolution studies provide a key consistency check for the mass
estimations assuming hydrostatic equilibrium. The presence of
radio sources within a cluster of galaxies is clearly a source of
disturbance for SZ measurements. The possibility to remove their
effect depends on the angular resolution of SZ observations. Radio
emission studies of some clusters reveal the presence of magnetic
fields and cosmic rays which are of fundamental importance to
understand the energy budget of such galaxy clusters. About 10\%
of the galaxy cluster are so called ``radio loud''. They show
diffuse synchrotron emission arising from halos, relics and
mini-halos.

Early high resolution (i.e. beam$\simeq\!13$'') SZ studies made
with the Nobeyama telescope have for instance allowed us to
understand that previously hypothesized relaxed clusters are,
indeed, into a merging activity \citep{kom01}. The same RX
J1347.5-1145 cluster was also observed by MUSTANG from GBT
(beam$\simeq\!9$'') \citep{mas10}, CARMA (synthesized
beam$\simeq\!11" \times 17"$) \citep{pla13} and NIKA from IRAM
30-m(beam$\simeq\!18.5"$) \citep{ada15}. \cite{fer11} found a correspondence between the mini-halos
emission and the high angular resolution SZ emission detected with
the MUSTANG experiment mounted on the GBT \citep{dic08} which can be
explained as gas oscillations within the gravitational potential
(i.e. sloshing \citep{maz08}). BOLOCAM \citep{gle98} and MUSTANG
observations have studied the galaxy cluster pressure profiles in
Abell 1835 and MACS-0647 \citep{rom15} while \cite{say13} performed an analysis over 45 massive clusters
cross correlating SZ and X-ray measurements and studying in
details relaxed, disturbed and cool-core clusters. Such
observations are in fact enabling, among the rest, to begin
understanding the galaxy cluster pressure profiles, the
mass-Comptonization scale relations, and to start investigating
the thermal state of the ICM in the outskirts of galaxy clusters.
Nevertheless, the picture is not complete. Both efforts, NIKA on
IRAM and MUSTANG on the GBT are being upgraded with
NIKA2 \citep{cal16} and MUSTANG2 \citep{dic14} instruments allowing to
enlarge the focal planes and the fields of view (f.o.v.)
reducing thus the need to rely on
larger-f.o.v./lower-angular-resolution experiments in order to
retrieve the large angular scales in their maps. Of clear interest
is also the effort spent by the AMI and CARMA Collaborations to
study the radio/microwave emission of galaxy clusters with
resolutions that can get to $\simeq\!0.3$\,arcmin \citep{ami13}. A
key opportunity is also represented by the newly built 64\,m
Sardinia radio telescope \citep{dam11} which, thanks to its primary
reflector active surface, provides a good efficiency up to
100\,GHz and is now opening to the scientific
community.\footnote{http:/\!/www.srt.inaf.it/media/uploads/astronomers/avpaperi.v4b.pdf.}

\section{SZ as Speedometer}

As mentioned in Sec.~\ref{par:ksz}, the kinematic SZ effect is in principle a
powerful method to retrieve the peculiar velocity, along the line of sight, of the
electrons of the ICM. This is important both to study the dynamics within galaxy
clusters with high angular resolution observations, for instance, in the violent
merger cases, and to measure the overall gas velocity in a highly complementary
way with respect to future high-dispersion X-ray spectroscopic observations such
as those planned by the $Athena+$ mission \citep{nan13}.

Millimetric/multifrequency observations are needed to disentangle the thermal SZ
and the kinematic SZ effects for a direct detection of the kinematic effect
itself. First pioneeristic millimetric observations of the SZ effect have been
made by MITO \citep{dep02} and SuZIE \citep{hol97} experiments. SuZIE II experiment
gave the first attempt to determine peculiar velocities on six galaxy clusters
providing, upper limits \citep{ben03}.

The first evidence of the motions of galaxy clusters through their kinematic SZ
signal has been provided using the 148\,GHz maps of the ACT experiment,
cross-correlating it with the baryon oscillation spectroscopic survey (BOSS) in
the Sloan Digital Sky Survey III \citep{KSZ3}. This is a statistical detection which
relies on a mean pairwise momentum estimation on the microwave maps, using the
BOSS catalog as galaxy clusters proxies. On the other hand, direct incredibly high
line-of-sight velocity of $v_{p}=-3450\pm900\,{\rm km/s}$ was reported by the
BOLOCAM Collaboration using data at 140\,GHz and 270\,GHz \citep{say13}.

A full investigation of the galaxy clusters motions is clearly still at the
primordial stage. Difficulty in disentangling the thermal and kinematic SZ effect,
and the latter from other sources of disturb is clearly hampering the full
exploitation of this methodology. In addition, the mixing of different effects,
including the relativistic corrections, makes it even harder.

\subsection{On scattering of CMB radiation on wormholes$\/:$ Kinematic SZ-effect}

Wormholes as candidate for dark matter particles have also been
studied \citep{KS07,KS11,KT}. The final choice requires the direct
observation of effects related to wormholes. We suggest to use the
kinematic SZ effect \citep{ZS,ZSa} which is known to have the
universal nature and is long used to study peculiar motions of
clusters and groups, e.g. \cite{KSZ1,KSZ2,KSZ3} (for details see
\cite{S151}). In the present section, we consider the
scattering of CMB radiation on wormholes and show that wormholes
also contribute to kinematic SZ effect and can be observed in
voids. Stable cosmological wormholes have throat sections in the
form of tori \citep{S15}. Upon averaging over orientations such tori
can be considered as spheres, and therefore, we use for estimates
the spherically symmetric model for a wormhole \citep{KSWS}.

The scattering of signals on spherical wormholes has been
previously studied in \cite{KSZ}, \cite{KSS},
and~\cite{KSWS} Consider first the case of a static gas of
wormholes, i.e. in the absence of peculiar motions. The spherical
wormhole can be considered as a couple of conjugated spherical
mirrors. When a relict photon falls on one mirror, a reflected
photon is emitted, upon the scattering, from the second
(conjugated) mirror. The cross-section of such a process has been
derived in \cite{KSWS} and can be summarized as follows.
Let an incident plane wave (a set of photons) falls on one throat.
Then, the scattered signal has two parts. First part represents
the standard diffraction (which corresponds to the absorption of
CMB photons on the throat) and forms a very narrow beam along the
direction of the propagation. This is the so-called scattering
forward which is described by the cross-section
\begin{equation*}
\frac{d\sigma _{absor}}{d\Omega }=\sigma _{0}\frac{\left(
ka\right) ^{2}}{4\pi }\left\vert \frac{2J_{1}\left( ka\sin \chi
\right) }{ka\sin \chi }\right\vert ^{2},
\end{equation*}%
where $\sigma _{0}=\pi a^{2}$, $a$ is the radius of the throat, $k$ is the wave
vector, and $\chi $ is the angle from the direction of propagation of the incident
photons, and $J_{1}$ is the Bessel function. Together with this part the second
throat emits an omnidirectional isotropic flux with the cross-section
\begin{equation}
\frac{d\sigma _{emit}}{d\Omega }=\sigma _{0}\frac{1}{4\pi }.
\label{flux}
\end{equation}%
It is easy to check that the total cross-sections coincide
\begin{equation*}
\int \frac{d\sigma _{absor}}{d\Omega }d\Omega =\int \frac{d\sigma _{emit}}{d\Omega }%
d\Omega =\sigma _{0};
\end{equation*}%
which is equivalent to a conservation law for the number of photons (the number of
absorbed and emitted photons coincides). This is enough to understand what is
going on with CMB in the presence of the gas of wormholes. In the absence of
peculiar motions, every end of a wormhole throat absorbs photons as the absolutely
black body, while the second throat end re-radiates them in an isotropic manner
(with the black body spectrum). It is clear that there will not appear any
distortion of the CMB spectrum at all. In other words, we may say that in the
absence of peculiar motions the distortion of the spectrum does not occur.

Consider now the presence of peculiar motions. The motion of one
wormhole throat end with respect to CMB causes the angle
dependence of the incident radiation with the temperature
\begin{equation*}
T_{1}=\frac{T_{\gamma }}{\sqrt{1-\beta _{1}^{2}}\left( 1+\beta _{1}\cos
\theta _{1}\right) }\simeq T_{\gamma }\left( 1-\beta _{1}\cos \theta
_{1}+...\right)
\end{equation*}%
where $\beta _{1}=V_{1}/c$ is the velocity of the throat end and $\beta_{1}\cos \theta _{1}=\left( \vec{\beta}_{1}\vec{n}\right) $, $\vec{n}$ is
the direction for incident photons. Therefore, the absorbed radiation has the spectrum%
\begin{equation*}
\rho \left( T_{1}\right) =\rho \left( T_{\gamma }\right) +\frac{d\rho \left(
T_{\gamma }\right) }{dT}\Delta T_{1}+\frac{1}{2}\frac{d^{2}\rho \left(
T_{\gamma }\right) }{dT^{2}}\Delta T_{1}^{2}+...,
\end{equation*}%
where $\rho (T_{\gamma})$ is the standard Planckian spectrum and $\Delta T_{1}(\beta
_{1}\cos \theta_{1})=T_{1}-T_{\gamma }$. It turns out that to the first-order in $\beta
_{1}$ such an anisotropy gives no contribution in the re-radiated photons and does not
contribute in the distortion of the spectrum. Indeed, in the reference system in which
the second end of the throat is at rest we have the isotropic flux (\ref{flux}) and,
therefore, integrating over the incident angle $\theta _{1}$, we find $\int\!\Delta
T(\cos \theta _{1}) d\Omega =0$. This means that to the first-order in $\beta _{1}$ the
second end of the throat radiates (in the rest reference system) the same blackbody
radiation with the same temperature $T_{\gamma }$. To next orders in $\beta _{1}$, a
nonvanishing contribution to the distortion of the spectrum $\rho (T_{1}) -\rho
(T_{\gamma})$ appears. However, to next orders, a more important feature will appear,
when we consider the actual wormhole sections in the form of tori. We consider such
details elsewhere.

Consider now the re-radiation of the absorbed CMB photons. In the first-order by $\beta
_{2}=V_{2}/c$ ($V_{2}$ is the peculiar velocity of the second end of the wormhole throat)
it radiates the black body radiation with the apparent surface temperature
(brightness)
\begin{equation}
T_{2}\simeq T_{\gamma }\left( 1+\beta _{2}\cos \theta _{2}+...\right)
\label{sbr}
\end{equation}%
where $\beta _{2}\cos \theta_{2}=( \beta_{2}m)$ and $m$ is the
unit vector pointing out to the observer. This is exactly the kinematic SZ effect.

Consider a collection (cloud) of wormhole throats. Chaotic peculiar motions of different
wormholes do not contribute to the CMB distortion (at least if wormholes are not too
big). We may expect to observe only coherent motions of the cloud. To obtain the net
energy--momentum transfer between the CMB radiation and the gas of wormholes, we have to
average over the wormhole, distribution. On average CMB photons undergo $\tau _{w}$
scatterings, where $\tau _{w}$ is the projected cloud optical depth due to the
scattering. If $n(r)$ is the number density of wormholes measured from the center of the
cloud, then $\tau _{w}$ is given by $ \tau _{w}=\pi \overline{a^{2}}\int\! n(r)d\ell $,
where the integration is taken along the line of sight and $
\overline{a^{2}}=\frac{1}{n}\int\! a^{2}n(r,a)da. $ Here $n(r,a)$ is the number density
of wormholes depending on the throat radius $a$. Since all wormhole throats have the
surface brightness (see Eq.~(\ref{sbr})) which is different from that of CMB, the
parameter $\tau _{w}$ defines (together with the peculiar velocity of the cloud $\beta
_2$) the surface brightness of the cloud itself.

\subsection{Polarization of the SZ effect}

Despite the unique possible applications of the kinematic SZ effect as galaxy
cluster speedometer, it is worth stressing once more that the use of the kinematic
SZ effect to retrieve galaxy clusters motions only provides an indication on the
electrons peculiar velocity along the line of sight. On the other hand, transverse
motions may in principle be studied through the SZ effect when its polarization
will be detected.

The SZ effect is expected to be polarized at levels proportional to powers of
$(v_{p}/c)$ and $\tau$. Polarization in the SZ effect mainly arises from four
different processes, all invoking the presence of a quadrupole:
\begin{itemize}
\item the first one is due to the radiation quadrupole due to the thermal SZ effect
itself, which is generated by a previous scattering elsewhere in the cores
of the local and nearby clusters. This is thus a double scattering process;
\item the second one is similar to the first one but it is due to the kinematic effect;
\item the third one is due to the radiation quadrupole due to Doppler shift caused by the
intrinsic peculiar velocity of the electron gas in the plane normal to the line of sight;
\item the fourth one is due to the scattering of the CMB intrinsic quadrupole and free
electrons in the cluster of galaxy.
\end{itemize}

The first polarization effect originates from the anisotropic optical depth to a given
location in the cluster. For example, toward the outskirts of a cluster one expects to
see concentric (radial) patterns of the linear polarization at frequencies where the
thermal SZ effect is positive (negative). It is thus due to the anisotropy in the
radiation field in the direction of the cluster center due to the thermal SZ effect
itself. Nevertheless, nonspherical morphology for the electron distributions will lead to
considerably complicated polarization patterns. The peak polarization of this signal will
be order of $\tau$ times the SZ effect signal, i.e. $\sim\!0.01 \Theta_{e}\tau^2$. In
principle, this effect could be used to measure the optical depth of the cluster and
therefore to separate $T_e$ and $\tau$ from a measurement of the thermal SZ effect.

In the case of double scattering originating from kinematic SZ
effect (second case), the effect is also sensitive to the
transverse ICM peculiar velocity and the intensity is expected to
be $\sim\!0.01\tau^2(v_{p}/c)$. Using $\tau = 0.01$ and a bulk
motion of $500$\,km\,s$^{-1}$, results in maximum polarization
levels of order few nK,  beyond the sensitivity of current
instrumentation.

The third case is unambiguously sensitive to the ICM velocity in the direction
transverse to the line of sight. In fact, in the case of polarization directly due
to the motion of the cluster with respect to the CMB, and transverse to our line
of sight, the quadrupole comes from the Doppler shift. This effect has been found
to be $\sim\!0.1\tau (v_{p}/c)^2$, again at the nK level even for hot
clusters \citep{ZS,Saz99,lav04}.

The fourth polarization possibility is due to Thomson scattering of the intrinsic CMB
quadrupole by the free electron of the ICM. \cite{Saz99}
calculated the expected polarization level and found the maximum CMB quadrupole
induced polarization is few $\tau \times 10^{-6}$K, somewhat higher than the expected
velocity induced terms discussed above. Even if it is still too small to expect detection
in the near future, this mechanism could possibly be used to trace the evolution of the
CMB quadrupole if polarization measurements could be obtained for a large number of
clusters binned in direction and redshift.

Despite in all cases the expected level of SZ polarization is lower than the
current instrumental sensitivity and the current ability to disentangle
cosmological effects from foregrounds, once the different SZ polarization will be
able to be disentangled, the SZ polarization will represent a unique possibility
to make 3D studies of galaxy cluster motions.

\section{Spectrometric Measurements of the SZ Effect}

Most of the effects described in this review can only be studied with
multifrequency detections of the SZ effect. All of them would greatly benefit from
spectroscopic measurements of the SZ with moderate spectral resolution (i.e.
5--10\,GHz). In addition, the capability to disentangle the SZ effect from
foregrounds, such as thermal dust emission, free--free and synchrotron, is
directly dependent by the number of spectral channels one makes the observations
with. Cosmological effects such as the deviation of the absolute CMB temperature
from a standard adiabatic scenario, detectable with the {\em M-R} method, show
their biggest effects in the ``cross-over'' band (i.e. $\sim\!217$\,GHz).
Astrophysical effects as the presence of nonthermal component caused by e.g. AGNs,
relativistic plasma, shock acceleration or dark matter (neutralinos) are predicted
to be important in the sub-mm region \citep{col03}. The possibility to disentangle
thermal SZ, kinematic SZ and its relativistic corrections, and to determine the
physical parameters of the galaxy clusters such as optical depth, plasma
temperature, peculiar velocity etc. requires detailed spectral informations in the
whole mm range. \cite{deb12} have shown that a
differential spectrometer covering four bands with a resolution of $\sim\!6$\,GHz
eliminates most of the bias resolution in a full recovery of the input cluster
parameters. This possibility is clearly hampered in ground-based few-band
photometers while spectrometers suitable for balloon borne observations or for
dedicated space missions would make the necessary step forward.

An early effort was produced by the Italian astrophysical community for the construction
of the Spectroscopic active galaxies and cluster explorer (SAGACE) experiment proposed to
the Italian space agency for a small mission Earth-orbit satellite with the intent to use
a 3\,m diameter primary mirror coupled to a Fourier transform spectrometer
(FTS) \citep{deb10}. More recently, the OLIMPO experiment \citep{olimpo} has been upgraded
with a differential FTS \citep{schillo,dal15} taking advantage of the clean conditions that
a balloon borne experiment can find in the stratosphere. OLIMPO will be launched on June
2017 from Svalbard Islands and will represent the first possibility to explore the use of
a FTS coupled with a four-band photometer in the millimetric and sub-millimetric range.
Simulations \citep{deb12} show that even a warm spectrometer, mounted on a balloon borne
experiment like OLIMPO \citep{olimpo}, allows an unbiased recovery of the galaxy cluster
thermal optical depth, electron temperature, dust contamination, non-thermal optical
depth, despite being still limited at high frequencies by radiative background. A low
earth-orbit satellite like SAGACE \citep{deb10} improves the capability to detect
non-thermal effects and the disentangling from foregrounds. A Sun-Earth Lagrangian point
L2 spectrometer like the one proposed for the Spectrum-M --- millimetron mission \citep{mm}
allows an exquisite recovery of the input parameters with better than 10\% precision with
no need to include strong priors on the input parameters.

This experimental activity is clearly connected to the renewed interest, in the
scientific cosmological community, to the possibility to detect the deviation from
a black body in the frequency spectrum of the CMB \citep{chluba} and the
experimental effort going on especially for the preparation of the next CMB space
mission for this goal \citep{pixie,prism}.

\section*{Acknowledgments}

The authors who are members of the {\it Planck} Collaboration warmly thank the
{\it Planck} Collaboration for numberless and constructive conversations on the
subjects discussed here. Some of the results in this paper have been derived using
the HEALPix \citep{Gor05} package. We acknowledge the use of the NASA Legacy Archive
for microwave Background Data Analysis (LAMBDA) and of the ESA {\it Planck} Legacy
Archive. C.~Burigana and M.~Rossetti acknowledge partial support by ASI/INAF
Agreement 2014-024-R.1 for the {\it Planck} LFI Activity of Phase E2. Some of the
results presented here are based on CFHT observations processed at the TERAPIX
data center located at IAP/France. We thank Gemma Luzzi for Fig.~2 on the $T_{\rm
CMB}$ versus $z$ scaling relation. We acknowledge the use of the SZ-Cluster
Database operated by the Integrated Data and Operation Center (IDOC) at the
Institut d'Astrophysique Spatiale (IAS) under contract with CNES and CNRS.

\end{document}